\documentclass[reprint,twocolumn,showpacs,preprintnumbers,amssymb,pre,superscriptaddress]{revtex4}
\usepackage{graphicx}
\usepackage{psfrag}
\usepackage{color}
\usepackage{amsbsy}
\usepackage{amssymb}       
\def\S{{\bf S}}
\def\dd{\mbox{d}}

\def\l{\ell}

\def\G{\Gamma}

\def\r{{\bf r}}

\begin{document}

$\:$

\title{Multistage adsorption of diffusing macromolecules and viruses}

\author{Tom Chou$^{1,2}$ and Maria R. D'Orsogna$^{2}$ \\ 
$^{1}$Dept. of Biomathematics, UCLA, Los Angeles, CA 90095-1766 \\
$^{2}$Dept. of Mathematics, UCLA, Los Angeles, CA 90095-1555}

\date{\today}

\begin{abstract}
We derive the equations that describe adsorption of diffusing
particles onto a surface followed by additional surface kinetic steps
before being transported across the interface.  Multistage surface
kinetics occurs during membrane protein insertion, cell signaling, and
the infection of cells by virus particles. For example, viral entry
into healthy cells is possible only after a series of receptor and
coreceptor binding events occur at the cellular surface.  We couple
the diffusion of particles in the bulk phase with the multistage
surface kinetics and derive an effective, integro-differential
boundary condition that contains a memory kernel embodying the delay
induced by the surface reactions. This boundary condition takes the
form of a singular perturbation problem in the limit where
particle-surface interactions are short-ranged. Moreover, depending on
the surface kinetics, the delay kernel induces a nonmonotonic,
transient replenishment of the bulk particle concentration near the
interface. The approach generalizes that of Ward and Tordai
\cite{WARDTORDAI} and Diamant and Andelman \cite{DIAMANT} to include
surface kinetics, giving rise to qualitatively new behaviors. Our
analysis suggests a simple scheme by which stochastic surface
reactions may be coupled to deterministic bulk diffusion.

\end{abstract}

\maketitle

\section{Introduction}

The kinetics of surface particle adsorption and of transport through
interfaces play a key role in surfactant phenomena \citep{BLANK,MALD},
membrane biology and cell signaling
\citep{BERG,RADKE,YBERT,SCHURR,SCHURR2,LIGANDREVIEW}, marine layer
oceanography \citep{SEA}, and other biological and chemical processes.
Particle adsorption may fundamentally alter the physical and chemical
properties of the interface, and it is crucial to understand both
equilibrium and dynamical properties of the adsorbed layers
\citep{WARDTORDAI,BLANK,MALD,RADKE}. In the seminal work of Ward and
Tordai \citep{WARDTORDAI}, a bulk phase acting as a reservoir of
particles is physically limited by an empty surface onto which the
particles can adsorb.  Particles are assumed to lower their free
energy with respect to the bulk phase by irreversibly and
instantaneously adsorbing onto the interface.  Under these conditions,
the total concentration of adsorbed particles may be estimated in
relation to measurable interfacial properties, such as the dynamic
surface tension.  Several applications, extensions and alternate
approaches to this work have been proposed \citep{DIAMANT,EDWARDS}. In
particular, adsorption dynamics in the Ward-Tordai setting can be
rederived through a free energy approach \citep{DIAMANT}, allowing for
the inclusion of ionic surfactant effects and electrostatic
interactions.

In many biochemical systems, the complete adsorption of a particle
arriving from the bulk requires a series of auxiliary transformations
at the surface before the particle can be successfully incorporated,
or `fused' into the surface. These intermediate steps gives rise to a
lag-time in the complete adsorption process.  For example, the
incorporation of emulsifying proteins onto an air-water interface may
be delayed by the unfolding of the polypeptide at the interface
\citep{YBERT}.  Adsorption of proteins on polymer-grafted interfaces,
such as the glycocalyx layer of vascular endothelial cells, is also
delayed due to the progressive insertion of the protein through the
polymer brush\cite{PURDUE2,PURDUE} . Kinetic delays have also been
observed in the adsorption of the hemagglutinin glycoprotein (HA) of
the influenza virus as it enters target host cellular membranes
\citep{HADELAY}.  The mechanisms underlying this delay are not known
in detail but are believed to involve conformational changes of HA
molecules into fusion enabling complexes, mediated by the presence of
binding receptors and coreceptors on the target cell membrane
\citep{HADELAY,HADELAY1,HADELAY2}.  Similarly, the incorporation of an
HIV particle into a T-cell or a macrophage is possible only after the
gp120 glycoprotein of the HIV virus membrane recognizes and binds to
the target cell surface receptor CD4, and subsequently to other
coreceptors such as CCR5 or CXCR4.  As in the case of HA and
influenza, the exact number of gp120-bound receptors and coreceptors
required for HIV particle fusion is yet unknown and might depend on
gp120 conformations and receptor/coreceptor binding cooperativity
\citep{KABAT1,KABAT2}.  The complex nature of surface biochemistry
makes quantitative kinetic measurements challenging. Recently, the
binding kinetics of the CD4 cellular receptor to the gp120 HIV ligand
have been measured under different experimental conditions yielding
widely different dissociation rates \citep{gp120,gp120b}. In this
work, we will provide a quantitative framework that can be used to
better understand the experimentally observed lag-times in surface
kinetics phenomena that involve multistage surface chemistry.

\begin{figure}[t]
\begin{center}
\includegraphics[height=2.1in]{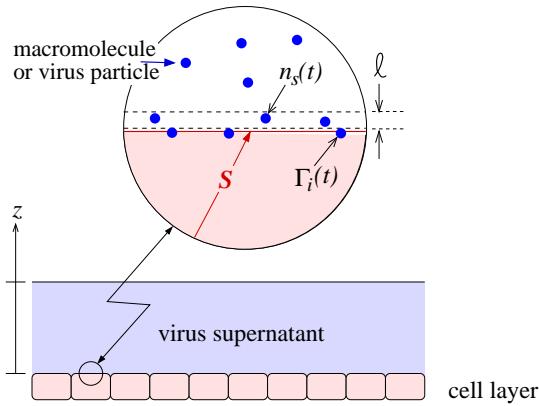}
\end{center} 
\caption{A schematic of a typical adsorption
  experiment on a confluent monolayer of target cells.  The location
  of the interface is labelled by ${\bf S}$. Particles such as viruses
  are spread over the cell layer in a thin supernatant film.  Inset:
  After initial nonspecific viral adsorption on the supernatant-cell
  interface, cellular receptors and coreceptors bind to the virus via
  a certain stoichiometry, forming fusion intermediates
  $\Gamma_i$. The subsurface layer of thickness $\l$, the subsurface
  concentration $n_{s}(t)$, and the adsorbed species $\G_{i}(t)$ are
  discussed in the text.}
\label{FIG1} 
\end{figure}

In particular, we will explicitly consider intermediate, reversible
steps for surface binding in the Ward-Tordai formalism, deriving an
effective boundary condition to complement the bulk diffusion
process. Chemical transitions among the surface species will introduce
memory terms in the boundary conditions for the bulk concentration.
Our analysis can be readily applied to the titration of
replication-incompetent virus via a colony formation assay
\cite{ANDERSON} as shown in Fig.\,\ref{FIG1}.

\section{Model Equations}

In this section, we motivate and derive the equations coupling bulk
diffusion to surface layer evolution. We consider a general, linear
reaction scheme to describe the multistep surface reaction dynamics.
Effective boundary conditions for diffusion from the bulk are derived
in Sect.\,3. As we shall discuss in detail, we are able to embody the
response of the adsorbing particle system to the existence of
intermediate chemical steps at the surface, into a unique delay kernel
regulating the boundary dynamics.  All microscopic details stemming
from the surface dynamics, no matter how complicated, are contained in
the derived memory kernel. Our approach includes ligand rebinding to
surfaces, found to be important for analyzing surface plasmon
resonance assays of biochemical systems \cite{TAUBER}.  In Sect.\,4 we
particularize our surface reaction scheme to a specific Markov process
chain and evaluate all physically relevant quantities.

\subsection{Bulk Diffusion}

In the continuum limit, the density of particles $n(\r,t)$ in the bulk
phase obeys the convection-diffusion equation

\begin{equation}
{\partial n \over \partial t} = \nabla\cdot\left[D \nabla
n\right] + \frac 1 {k_B T}\nabla\cdot\left[{D}\,n\nabla U \right],
\label{CONVECTIONDIFFUSION}
\end{equation}


\noindent where $D(\r)$ and $U(\r)$ are the local diffusion
coefficient and potential of mean force, respectively, and $k_{B}T$ is
the thermal energy. Spatial variation of $D(\r)$ and $U(\r)$ may arise
from interactions with the interface as shown in Fig.\,\ref{LAYER}.
Boundary conditions are typically applied at the mathematical surface
onto which the particles adsorb or reflect. By balancing the diffusive flux
just above  this mathematical interface with the particle
rate of insertion into the 
interface, a mixed boundary condition arises 

\begin{equation}
D(\r)\,{\bf \hat n}\cdot\nabla n(\r,t) =  \gamma n(\r,t), \quad \r\in {\bf S}.
\label{BC0}
\end{equation}

\noindent Here, ${\bf S}$ denotes the substrate; its normal direction
is ${\bf \hat n}$.  The parameter $\gamma$, which has the physical
units of speed, is proportional the probability $f$ (often called the
accommodation coefficient \cite{LANDAU} or sticking probability
\cite{JUNG,KISLIUK}) that a particle is adsorbed into the mathematical
interface upon collision. We define $\gamma = \gamma_{0}f$ such that
in the limit $\gamma_0 \rightarrow \infty$ and $f\neq 0$,
Eq.\,\ref{BC0} is equivalent to $n(\r \in \S) = 0$, an absorbing
boundary condition.  A reflecting boundary condition, ${\bf \hat
n}\cdot\nabla n=0$, arises when $f = 0$.  Equations
\ref{CONVECTIONDIFFUSION} and \ref{BC0} are commonly used to model
simple diffusion-adsorption processes at surfaces.

\subsection{Surface Reactions}

In many applications, particles at an interface undergo chemical or
physical modifications that control for example, surface reactivity,
surface tension \cite{WARDTORDAI,DIAMANT,BLANK,MALD}, and conductivity
\cite{JUNG}. Biological examples include tissue factor initiated
coagulation reactions and viral entry. Coagulation factors must work
their way through the glycocalyx layer before they can be
enzymatically primed by the membrane-bound tissue factors
\cite{MCGEE}.  Entry of viruses, such as HIV, into cells require the
binding of membrane-bound receptors and coreceptors before fusion with
the target cell can occur. All of these processes can be thought of as
reactions at the membrane surface.  Immediately after adsorption from
the bulk, the surface particle concentration, whether of coagulation
factors or of virus particles, is denoted by $\G_{1}$. For example, in the
case of viruses, we can identify the $\G_1$ state as being that of a
virus bound to $i=1$ CD4 surface receptor.  The initially adsorbed
species can then kinetically evolve into the other species $\G_{i}$
representing virus particles with $i >1 $ bound receptors or
coreceptors. The kinetics among the $N$ surface species follows the linear
rate equation $\partial_{t}{\bf \G} = {\bf M}{\bf \G} + {\bf F}(t)$,
where ${\bf \G}\equiv (\G_{1},\G_{2},\ldots,\G_{N})$, ${\bf M}$ is the
transition matrix among the $N$ surface states, and ${\bf F} \equiv
(F,0,\ldots,0)$ is the source of the first, originating source species
$\G_{1}$ coming from the bulk.

\vspace{3mm}

\begin{figure}[t]
\begin{center}
\includegraphics[height=0.35in]{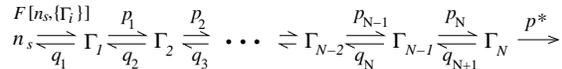}
\end{center} 
\caption{Surface reaction scheme among the
intermediate states $\G_{i}$.  The source species $\G_{1}$ is supplied
by the bulk surface concentration at the interface, $n_{s}$.  In the
context of virus recognition and infection, the intermediate steps
label various numbers of receptors or coreceptors associated with the
surface bound virus particle. For example, $\G_{i}$ may denote the
surface concentration of HIV particles with $i$ receptors and
coreceptors attached. This catenary model could also represent
successive degrees of insertion of an absorbing species through a
polymer brush or glycocalyx coated interface.}
\label{CHAIN} 
\end{figure}
Figure \ref{CHAIN} illustrates a simple example of
a linear surface reaction scheme that can be described by the above
linear rate equation. In this case the reaction matrix ${\bf M}$ is
tridiagonal.  General reaction matrices can also be analyzed since our
results depend only on the eigenvalues and eigenvectors of
${\bf M}$.

\subsection{Surface Layer}

Because the surface densities $\Gamma_{i}$ carry units of number per
area, and the bulk densities are expressed by number per volume, any
kinetic parameter linking bulk source concentrations to those at the
interface must introduce a physical length scale.  Diamant and
Andelman \cite{DIAMANT} have introduced the sublayer thickness as a
mathematical step coupling the bulk density to surface density.  Here,
we physically motivate this ``surface layer'' and the associated
transport.  Let us thus introduce a thin layer of thickness $\l$ near
the surface, in which the particle density is denoted $n_{s}(t)$ and
is still expressed in units of number per volume.

\begin{figure}[t]
\begin{center}
\includegraphics[height=2.0in]{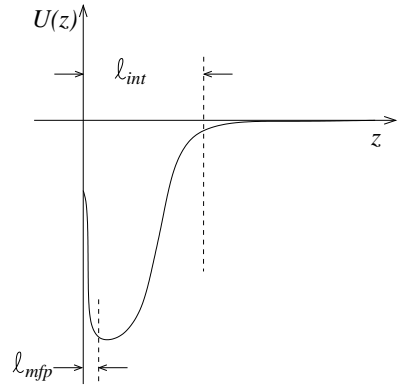}
\end{center} 
\caption{
Interaction potential between diffusing species and the
interface.  The subsurface layer is defined by either the range of the
interaction potential $\l_{int}$, or the mean free path $\l_{mfp}$,
depending upon which is larger.}
\label{LAYER} 
\end{figure}

The continuum approximation Eq.\,\ref{CONVECTIONDIFFUSION} breaks down
when resolving the transport within distances of a few mean free
paths.  If we identify the sublayer thickness $\l$ with the mean free
path $\l_{mfp}$, as shown in Fig.\,\ref{LAYER}, we must solve
Eq.\,\ref{CONVECTIONDIFFUSION} with a nonuniform $U(\r)$, and possibly
a nonuniform $D(\r)$, to within a distance $\l = \l_{mfp}$ of the
interface. For the choice $\l=\l_{mfp}$, the adsorption velocity
$\gamma_{0}$ can be approximated by the thermal velocity $v_{T}$ such
that $\gamma=\gamma_{0}f \sim v_{T}f$. The value of the bulk density
at the boundary $\S + \l {\bf \hat n}$ is defined as the sublayer
density: $n(\r=\S+\l{\bf \hat n},t) \equiv n_{s}(t)$. The equation for
the rate of change of the number per area of molecules in the thin
layer, $d(\l n_{s}(t))/dt$ can be obtained by balancing the latter
with the diffusive flux into the layer $D(\r)\,{\bf \hat n}\cdot\nabla
n(\S + \l{\bf \hat n},t)$, the adsorption into the surface
concentration $\G_{1}$, and the spontaneous desorption from the
initially adsorbed species $\G_{1}$ occurring at rate $q_{1}$.  The
complete set of equations coupling the variables $n(\r,t), n_{s}(t),$
and $\G_{i}$ is thus

\begin{eqnarray}
{\partial n \over \partial t} &=& \nabla\cdot\left[D \, \nabla
n\right] + \frac 1 {k_B T} \nabla\cdot\left[{D}\,n \, \nabla U \right],
\label{CONVECTIONDIFFUSION2} \\
\displaystyle \l{\dd n_{s} \over \dd t} &=& -F + D \,{\bf \hat
n}\cdot\nabla n \, \Big|_{\r = \S+\l{\bf n}} + q_{1}\, \G_{1},\label{NS} \\
{\dd {\bf \G} \over \dd t} &=& 
{\bf M}{\bf \G} + {\bf F}, \quad \, {\bf F}
=(F,0,0,\ldots,0).
\label{GAMMA}
\end{eqnarray}

\noindent Here, $n |_{\r = \S + \l{\bf n}} = n_{s}$, and $F=F[n_{s},
\{\G_{i}\}]$ is the flux of the surface concentration $n_{s}$ into the
incipiently adsorbed species $\G_{1}$.  This functional may depend on
interactions among the adsorbed species $\Gamma_{i}$, including
cooperative or crowding effects, and may be modeled using free
energies and chemical potential differences between the bulk and
surface \cite{DIAMANT,PURDUE2}.

A further simplification can be introduced by defining a different
sublayer thickness $\l = \l_{int}$, where $\ell_{int}$ is the typical
range of the particle-surface interaction as shown in
Fig.\,\ref{LAYER}.  In this case, at least at a
distance $\ell_{int}$ from the interface, 
$D(\r)$ is constant, $U(\r)$ is zero and
Eq.\,\ref{CONVECTIONDIFFUSION} is approximated by the
standard diffusion equation

\begin{equation}
{\partial n(\r, t)\over \partial t} = D\nabla^{2}n(\r,t).
\label{DIFFUSION0}
\end{equation}

All effects of the potential of mean force $U(\r)$ and spatially
varying $D(\r)$ are now subsumed into an effective source $F[n_{s},
\{\G_{i}\}]$. This is consistent with all previous treatments
\citep{WARDTORDAI,DIAMANT,MALD} in which transport in the bulk phase
was described by simple diffusion with uniform $D$ and $U$.  Provided
$F[n_{s}, \Gamma_i]$ is independent of $\G_{i}$, equations
\ref{CONVECTIONDIFFUSION}-\ref{GAMMA} can be explicitly solved in
simple geometries. For low surface densities $\G_{i}$ such that
additional adsorption is not hindered by steric exclusion, one can
assume a form $F[n_{s}]$ independent of surface concentrations
$\Gamma_{i}$.  For $\ell=\l_{int}$, $\gamma$ is now interpreted as an
effective adsorption coefficient allowing
Eq.\,\ref{CONVECTIONDIFFUSION} to be replaced by
Eq.\,\ref{DIFFUSION0}, and simplifying the bulk concentration
equation.

\section{Analysis}

Following the original work of Ward and Tordai, subsequent studies on
adsorption and dynamic surface tension measurements
\cite{WARDTORDAI,MALD} eliminate the bulk density at the interface in
Eq.\,\ref{NS} to yield two coupled integro-differential equations for
$n_{s}$ and $\G_{1}$ which must be numerically self-consistently
solved.  Here, we solve the linear Eqs.\,\ref{GAMMA} independently
from the bulk densities, but with a source term $F[n_{s}]$ that
connects the surface concentrations $\G_{i}$ with the bulk
concentration $n(\r,t)$. If $F[n_{s}]$ is independent of $\G_{i}$, the
explicit solution to Eqs. \ref{GAMMA} can be found by evaluating the
eigenvalues $\lambda_{j}$ and corresponding eigenvectors ${\bf v}^{j}$
of the chemical transition matrix ${\bf M}$. Denoting the similarity
transform matrix $V_{kj} \equiv v_{k}^{j}$ such that ${\bf V}{\bf
M}{\bf V}^{-1} =$diag($\lambda_{j}$), the surface densities are

\begin{equation}
\begin{array}{rl}
\Gamma_{k}(t)& \displaystyle = \sum_{j,m = 1}^{N} V^{-1}_{kj}V_{jm}
\Gamma_{m}(0)e^{\lambda_{j}t} \\[13pt]
\:  & \hspace{6mm} \displaystyle +
\sum_{j=1}^{N}V^{-1}_{kj}V_{j1}\int_{0}^{t}\!e^{\lambda_{j}(t-t')}F[n_{s}(t')]\dd t',
\label{GAMMASOLN}
\end{array}
\end{equation}

\noindent where ${\bf \Gamma}(0)$ are the intermediate surface
concentrations at $t=0$.  If there are no spontaneous sources of the
surface cell intermediates, all eigenvalues $\lambda_j < 0$.  From
Eq.\,\ref{GAMMASOLN}, in the case ${\bf \G}(0)=0$, ${\bf \G}(t)$ is
proportional to $F[n_{s}(t)]$. Upon substituting 
$\G_{1}(t)$ 

By setting $k=1$ in Eq. \ref{GAMMASOLN}, we substitute $\G_{1}(t)$
into Eq.\,\ref{NS}, and find a concise description of the
diffusion-adsorption process:

\begin{equation}
{\partial n(\r, t)\over \partial t} = D\nabla^{2}n(\r,t), \quad \quad 
n(\S+\l{\bf n},t) = n_{s}(t),
\label{DIFFUSION1}
\end{equation}

\begin{equation}
\l {\dd n_{s}(t) \over \dd t} = 
D {\bf \hat n}\cdot\nabla n(\r,t) \Big|_{\r={\bf S}+\l{\bf n}}
\!\! - \int_0^{t}\!K(t-t') F[n_{s}(t')] \dd t',
\label{EFFECTIVEBOUNDARY}
\end{equation}

\begin{equation}
K(t) = \delta(t) - q_{1}\sum_{j=1}^{N}V^{-1}_{1j}V_{j1}e^{\lambda_{j}t},
\label{KERNEL}
\end{equation}

\noindent where $K(t)$ is the kernel constructed from the eigenvalues
and eigenvectors of ${\bf M}$.  It is composed of an instantaneous
response -- the immediate depletion of $n_{s}$ due to adsorption into
the $\G_{1}$ surface species -- and delay terms arising from the
surface kinetics of Eq.\,\ref{GAMMA}.  The complete set of equations
\ref{DIFFUSION1}-\ref{KERNEL} is one of our main findings. This result
explicitly shows how multistage adsorption is modeled by a bulk
diffusion equation with an nonlinear integro-differential boundary
condition that incorporates the delay arising from the multistep
kinetics. Under this scheme, all effects of surface reactions are
incorporated in the kernel $K(t)$.

Our analysis can be carried further by specifying a linear form for
the $\Gamma_{i}$-independent source term

\begin{equation}
F[n_{s}(t)] = \gamma n_{s}(t),
\label{LINEARAPPROX}
\end{equation}

\noindent which simply takes the source for the surface concentration
$\Gamma_{1}$ of the first species to be proportional to the subsurface
concentration.  The surfaces densities $\G_{i}(t)$ can be found by
substituting $n_{s}(t)$, derived from
Eq.\,\ref{DIFFUSION1}-\ref{KERNEL}, into the expression for
$F[n_{s}(t)]$ in Eq.\,\ref{GAMMASOLN}. Note that the boundary
condition Eq.\,\ref{EFFECTIVEBOUNDARY} contains a singular
perturbation, and that for times $t \gg \l/\gamma$, the ``outer
solution'' approximation $\l (\dd n_{s}/\dd t) \approx 0$ yields the
standard mixed boundary condition Eq.\,\ref{BC0} with an additional
memory kernel. Moreover, in the linear approximation of
Eq. \ref{LINEARAPPROX}, the convolution of the delay term in the
effective boundary condition \ref{EFFECTIVEBOUNDARY} is amenable to
analysis by Laplace transforms. 

For simplicity, we will assume a simple one-dimensional problem where
all quantities vary spatially only in the direction normal to an
infinite, flat interface at $z=0$. We nondimensionalize all quantities
by using $\l$ as the unit of length, and $q_{1}^{-1}$ as the unit of
time. Henceforth, in all equations, we make the replacements
$z\rightarrow \bar {z}/\ell, \, t \rightarrow q_{1} \bar{t}, \,
n\rightarrow \l^{3} \bar {n}, \, n_{s}\rightarrow \l^{3} \bar
{n}_{s},\, \G_{i} \rightarrow \l^{2}\ \bar{G}_{i},\, D\rightarrow
\l^{-2} \bar{D}/q_{1},$ and $\gamma \rightarrow \bar{\gamma} /(\l
q_{1})$. To render the notation less cumbersome we omit the bars from
the redefined quantities. In the discussion that follows, the
$z,t,n,n_s,\Gamma_i,D, \gamma$ parameters are intended as
nondimensional.  Upon taking the Laplace transform in time of the
dimensionless forms of Eqs.\,\ref{DIFFUSION1},
\ref{EFFECTIVEBOUNDARY}, and \ref{KERNEL}, we obtain

\begin{equation}
s\tilde{n}(z,s)-n_{0}= D\partial_{z}^{2}\tilde{n}(z,s),
\label{LPDIFFUSION1}
\end{equation}

\begin{equation}
s\tilde{n}_{s}(s)-n_{0}=D\partial_{z}\tilde{n}(z,s)\Big|_{z=1}-
\gamma \tilde{K}(s)\tilde{n}_{s}(s),
\label{LPNS}
\end{equation}

\noindent where 

\begin{equation}
\tilde{K}(s) = 1-\sum_{j=1}^{N}{V^{-1}_{1j}V_{j1} \over s-\lambda_{j}},
\label{KERNELS}
\end{equation}

\noindent and $n_{0}$ is the initial, dimensionless constant bulk and
sublayer concentration. The general solution to the bulk density
$\tilde n(z,s)$ from Eq.\,\ref{LPDIFFUSION1} is

\begin{eqnarray}
{\tilde{n}(z,s)\over n_{0}} = {1\over s} - {\gamma
\tilde{K}(s)\exp\left(-(z-1)\sqrt{s/D}\right)\over s(s+\sqrt{sD}+\gamma
\tilde{K}(s))}.
\label{NSZ}
\end{eqnarray}

\noindent 
Once the bulk density is derived, all other quantities can be found by
inverse Laplace transforming $\tilde{n}(z,s)$.  In the absence of
spontaneous sources of the surface intermediates, $\lambda_{j} <
0$. In this case, it is possible to show that $\tilde{n}(s,z)$ only
has a simple pole at $s=0$ and a branch cut on
$s=(-\infty,0]$. Performing the integral along the latter, we find the
exact results

\begin{equation}
n(z,t) =  n_{0} \int_{0}^{\infty}L(z,u)e^{-ut} \dd u,
\label{NZT}
\end{equation}

\noindent and 

\begin{equation}
\Gamma_{k}(t) = \gamma n_{0} \,\sum_{j=1}^{N}V^{-1}_{kj}V_{j1}
\int_{0}^{\infty} {e^{\lambda_{j}t}-e^{-ut} \over u+\lambda_{j}} L(1,u) \dd u,
\label{GAMMA2}
\end{equation} 

\noindent where 

\begin{widetext}
\begin{eqnarray}
L(z,u) \equiv\displaystyle  -{\gamma \over \pi}{(u-\gamma
\tilde{K}(-u))\sin\sqrt{{u\over D}}(z-1) - \sqrt{uD}\cos\sqrt{{u\over D}}(z-1)
\over u(u-\gamma \tilde{K}(-u))^2 + Du^2}\tilde{K}(-u).
\label{L} \\
\nonumber
\end{eqnarray}
\end{widetext}

\noindent Equations \ref{NZT}-\ref{L} are used to numerically compute
all of our results in the next Section. For completeness, analytic
expressions for asymptotically short and long time limits are derived
in the Appendix.

\section{Results}

We now specify a surface reaction scheme and construct its delay
kernel by using its associated eigenvalues and eigenvectors.  For
applications such as multiple receptor binding of the adsorbed
species, we consider a reversible sequential Markov process among $N$
chemical intermediates $\G_{i},\, i=1,\ldots,N$. As shown in
Fig.\,\ref{CHAIN}, formation of state $\G_{i+1}$ from state $\G_{i}$
occurs at rate $p_{i}$, while the reverse step occurs at rate
$q_{i+1}$. The final state $\G_{N}$ is irreversibly annihilated, by
transport across the membrane, or by irreversible reaction, with rate
$p^{*}$.  We can then explicitly write the dimensional form 
of Eq.\,\ref{GAMMA}, where ${\bf M}$ is  a
tridiagonal transition matrix, as

\begin{equation}
\begin{array}{rcl}
\displaystyle {\dd \G_{1} \over \dd t} &=& \displaystyle F[n_{s},
\{\G_{i}\}] - (p_{1}+q_{1}) \G_{1}+q_{2}\G_{2}, \\[13pt] 
\displaystyle
{\dd \G_{i} \over \dd t} &=& \displaystyle
p_{i-1}\G_{i-1}-(q_{i}+p_{i})\G_{i}+q_{i+1}\G_{i+1} \quad 
2 \leq i \leq N-1 \\[13pt]
\displaystyle  {\dd \G_{N} \over \dd t} &=& 
\displaystyle -(p^{*}+q_{N})\G_{N}+p_{N-1}\G_{N-1}. \\
\nonumber
\end{array}
\end{equation}

\begin{figure*}[t]
\begin{center}
\includegraphics[height=2.75in]{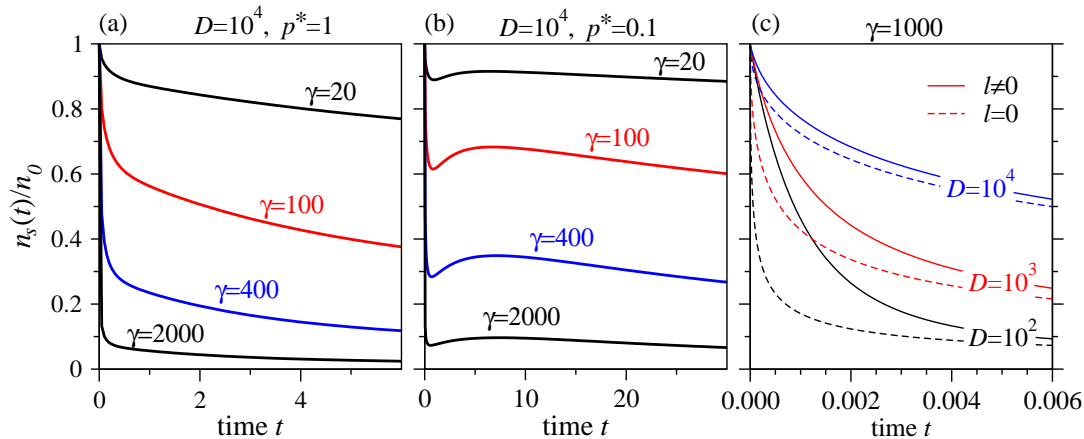}
\end{center} 
\vspace{-1cm}
\caption{Surface densities $n_s (t)$ for $N=1$. (a)
The sublayer density $n_{s}(t)$ as a function of time, for various values
of the dimensionless adsorption rate $\gamma$. The other parameters
are fixed at $D=10^{4}$ and $p^{*}=1$. (b) The sublayer density
$n_{s}(t)$ for $D=10^{4}$, and $p^* = 0.1$.  Note the bump in
concentration imparted by the slow annihilation rate $p^*$. (c) The
deviation of $n_{s}(t)$ (for $\gamma=1000$) if the transient term $\ell
dn_{s}/dt$ is neglected. The solid curves are found from the full
equation \ref{LPNS}, while the dashed curves are solutions when the
left-hand-side of Eq.\,\ref{LPNS} is neglected. The deviation occurs
only at very short times, is independent of $p^{*}$, and is greatest
for smaller diffusion coefficient $D$.}
\label{PLOT1} 
\end{figure*}

\noindent
In the simplest case where there is only one surface intermediate
before transport across the interface, $N=1$ and the dimensionless
($q_{1}=1$) delay kernel is simply $\tilde{K}(s) =
-(s+p^{*})/(s+p^{*}+1)$.  The sublayer concentration $n_{s}(t)$
derived from Eq.\,\ref{NZT}, and the surface concentration evaluated
from Eq.\,\ref{GAMMA2}, are shown in Figs.\,\ref{PLOT1} for various
values of $\gamma$.

Let us estimate typical parameter values for viral fusion or molecular
binding processes.  Typical diffusion constants for viruses of
diameter $100$nm and in aqueous environments, are $D \sim
10^{-8}$cm$^{2}$/s. Using the typical screened electrostatic
interaction potential, $\l \approx 10^{-7}$cm, we estimate the
dimensionless diffusion coefficient $D\sim 10^{6}$s$^{-1}/q_{1}$. On
the other hand, typical diameters of small ligand molecules are of the
order of 1nm yielding a nondimensional diffusion constant $D\sim
10^{8}$s$^{-1}/q_{1}$.  The nondimensional $\gamma$ estimated using
the thermal velocity $v_{T}$ is now $\gamma \approx f v_{T}/(\l
q_{1})$.  For virus particles $\gamma \sim 10^{8}$s$^{-1}f/q_{1}$,
while for molecular ligands $\gamma \sim 10^{10}$s$^{-1}f/q_{1}$. The
dissociation rate $q_{1}$ is highly variable and typically falls in
the broad range $q_{1} \sim 10^{-4}$ s$^{-1}$ to $10^{4}$s$^{-1}$.
For the gp120-CD4 interaction, the dissociation has been estimated in
model systems \cite{gp120} to be $q_{1} \sim 10^{-3}$s$^{-1}$, while
the detachment rate for mutant viral species \cite{gp120b} can be as
high as $q_{1} \sim 10^3$s$^{-1}$.  Lower dissociation rates are
possible in tighter binding ligand receptor pairs such as EGF-receptor
\cite{EGF} where $q_{1}\sim 10^{-4}$s$^{-1}$. For other pairs such as
P-selectin and its receptors \citep{PSELECTIN,PSELECTIN2}, $q_{1}\sim
0.1 $s$^{-1}$ to $1$ s$^{-1}$.  The sticking probability $f$ is
proportional to the binding probability of upon ligand-receptor
contact, multiplied by the receptor area fraction at the
interface. The factor $f$ depends on the receptor density, but is
typically of the order $f\sim 10^{-4} - 10^{-2}$.

In Fig.\,\ref{PLOT1}(a) we plot the sublayer density $n_s$ as a
function of time.  For $p^{*}=1$, Fig.\,\ref{PLOT1}(a) shows that the
sublayer density $n_{s}(t)$ starts at its initial value $n_{0}$ and
decreases with a nondimensional rate proportional to $\gamma$,
eventually monotonically reaching $n_{s}(t\rightarrow \infty)
\rightarrow 0$.  If the annihilation rate $p^{*}$ is decreased,
$n_{s}$ may no longer be monotonic. The observed increase in the
surface concentration is due to the slow consumption of material at
the interface, allowing some of the material to desorb after being
delayed at the interface, rather than irreversibly reaching the final
annihilated or fused state.  For example, when $p^*=0.1$, the surface
concentration $n_{s}(t)$ first decreases but recovers slightly at longer
times, before ultimately decaying to zero as shown in
Fig.\,\ref{PLOT1}(b).

\begin{figure}
\begin{center}
\includegraphics[height=4.3in]{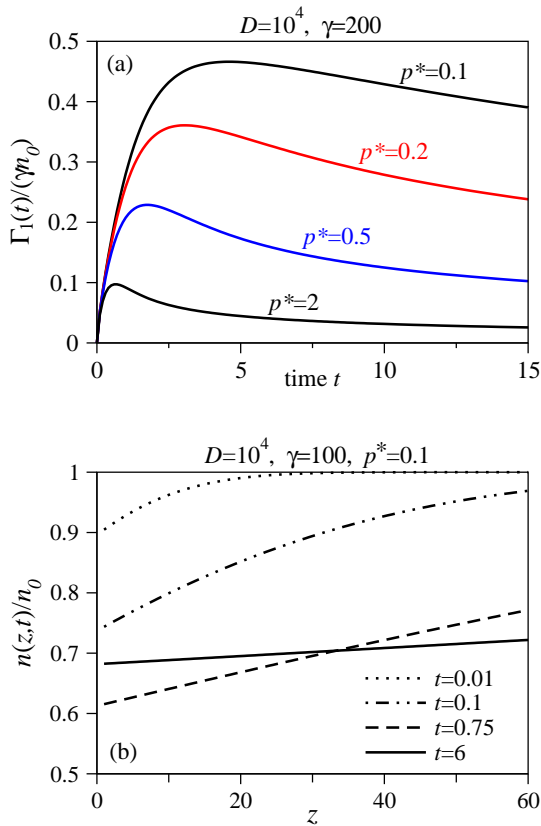}
\end{center} 
\vspace{-6mm}
\caption{
(a) The surface concentration $\Gamma_{1}(t)$
for various $p^*$ with $D=10^{4}$ and $\gamma=200$.  (b) The bulk
density profiles $n(z,t)$ as a function of position $z$ at times
$t=0.01,0.1, 0.75$, and $6$.}
\label{PLOT2} 
\end{figure}

Note that the curves for $n_{s}(t)$ exhibit a transient at short times
determined by $1/\gamma$. Beyond this transient, the full solution we
plot in Fig.\,\ref{PLOT1}(b) reduces to the outer solution,
corresponding to setting $\dd n_{s}/\dd t = 0$ on the left hand side
of Eq.\,\ref{EFFECTIVEBOUNDARY}. Fig.\,\ref{PLOT1}(c) explicitly shows
different behaviors of the full and outer solutions during the
transient time.  The effect of the $\ell dn_{s}/dt$ term is to slow
down the initial decrease of $n_{s}$, particularly for short times
within the transient defined by $1/\gamma$.  The effects of neglecting
the boundary layer are less pronounced for larger bulk diffusivities
$D$.

The temporal evolution of $n_s(t)$ is strongly dependent on
$\Gamma_1(t)$. In fact, the nonmonotonicity of $n_s(t)$ for small
$p^{*}$ shown in Fig.\,\ref{PLOT1}(b) arises from the build-up of
$\Gamma_{1}$ indicated in Fig.\,\ref{PLOT2}(a) which can get released
back into the subsurface layer. For smaller $p^{*}$, $\Gamma_{1}$
reaches larger values.  As long as $p^* > 0$, both $n_{s}$ and
$\Gamma_{1}$ vanish at sufficiently long times. Complete particle
depletion near the surface occurs in dimensions less than two because
there is no bounded steady-state solution to the diffusion equation
and the depletion zone moves away from the interface for all times as
shown in Fig.\,\ref{PLOT2}(b).  Despite free diffusion, the bulk is
unable to sustain a particle source near the surface as is known from
classic diffusion theory \cite{CRANK}.  The replenishment at small
annihilation rates $p^*$ also manifests itself in the bulk. In the
case shown in Fig.\,\ref{PLOT2}(b), as time increases from $t=0.75$ to
$t=6$, the bulk concentration near the interface recovers before
ultimately decreasing according to Eq.\,\ref{LONGTIME}.

\vspace{3mm}
\begin{figure}[t]
\begin{center}
\includegraphics[height=4.3in]{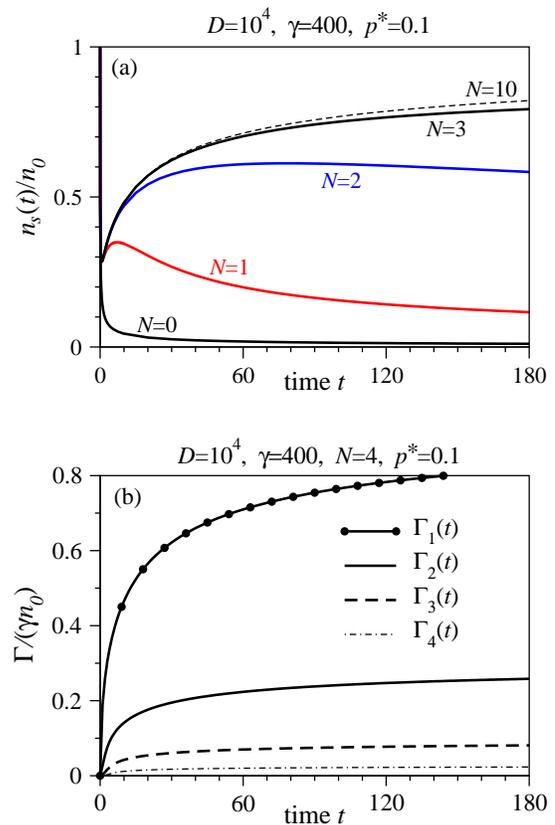}
\end{center} 
\caption{
Dependence of surface quantities on number of steps $N$ in
the surface reaction scheme. (a) The sublayer density $n_{s}$ as $N$
is increased. The initial rapid fall from $n_{s}/n_{0} = 1$ is
imperceptible on this scale. (b) The surface concentrations
$\Gamma_{i}$ as a function of time for $N=4$.}
\label{PLOT3} 
\end{figure}

For general $N$, the eigenvectors and eigenvalues must be explicitly
computed.  In Fig. \ref{PLOT3}(a), we plot $n_{s}$ as a function
of $N$ for uniform $q=1$ and uniform $p$. For small $p$, the surface
kinetics is a highly biased random walk away from $\Gamma_{N}$ toward
$\Gamma_{1}$, resulting in a larger $n_{s}$. Both small $p$ and large
$N$, hinder the annihilation process and impart a more reflective
character to the interface. After initial transients, both $n_{s}$ and
the surface concentrations $\Gamma_{i}$ maintain a high level for a
long time before dissipating. Larger $N$ also effectively trap surface
material in the surface reservoir $\Gamma_{i}$.  The relative amounts
of $\Gamma_{i}$ for $N=4$ are shown in Fig.\,\ref{PLOT3}(b). For the
small $p=0.1$ used, most of the surface density lies in the initial
species $\Gamma_{1}$, decreasing in the latter species.

\section{Summary and Conclusions}

We derived an effective, integro-differential equation for the
boundary condition of a simple diffusion process. The approach
presented differs from the typical Ward and Tordai treatments since we
use a linear model for the time rate of change of the initially
adsorbed species $\G_{1}$, rather than eliminating the bulk diffusion
equation.  The effects of intermediate chemical steps at the boundary
are described by a delay kernel that can be decomposed using Laplace
transforms. This kernel is an explicit function of the eigenvalues and
eigenvectors of the surface reaction transition matrix. Our results
suggest that measurement of a few quantities, such as fluorescence
monitoring of the sublayer density \cite{NLT}, can be used to
reconstruct the principal components of $K(t)$. This approach can be
used to probe qualitative features of the surface kinetics important
in modeling cell membrane signaling and viral infection, where a
sequence of chemical steps at the surface are required before
initiation of signaling or viral fusion. In HIV infection, the initial
adsorption rate would be proportional to the surface CD4
concentration, and the subsequent rates in the reaction scheme in
Fig.\,\ref{CHAIN} would depend on the coreceptor concentrations, their
surface mobilities, as well as the effects of cooperative binding
\cite{KABAT1}. All of these physical attributes are encoded in the
distribution of eigenvalues and eigenvectors of ${\bf M}$.

For simple linear reaction schemes on a flat surface, we find explicit
dependences of the surface concentrations $\Gamma_{i}$ and sublayer
concentrations on the eigenvalues and eigenvectors of the transition
matrix. For smaller annihilation rates $p^*$, and at least one ($N
\geq 1$) surface intermediate, we find that the surface concentration
persists and can replenish the bulk concentrations $n_{s}$ after its
initial decay.  The depletion zone in the bulk can also
recover. Delays that induce instabilities in dynamical systems have
been well established \cite{DELAY}. Here, although the delay occurs in
a boundary condition, we observe nonmonotonic behavior arising in the
bulk concentrations as well. This rebounding effect is also apparent
if one differentiates Eq.\,\ref{EFFECTIVEBOUNDARY} with respect to
time, giving a second order, harmonic oscillator-like equation, plus a
dissipative coupling to the bulk concentration.

Whether the surface concentration or the bulk concentration near the
surface vanishes at long times depends on the surface kinetics as well
as geometry. If the combined surface kinetics towards annihilation is
slow relative to bulk diffusion, the decay of the sublayer
concentration $n_{s}$ can be extremely slow. Similarly, if the number
of surface states is large, there is an effective delay to
annihilation and a higher probability that a surface species can
detach and replenish the sublayer concentration.  This effect is very
sensitive to the annihilation rate $p^*$ and the size of the reaction
$N$ and can keep the subsurface concentration high for essentially all
times.

A number of extensions and related approaches to this and related
systems can be readily investigated.  For example, in applications
such as surfactant adsorption, the surface concentration ${\bf \G}$
can be appreciable and suppress additional adsorption. If surface
species $\G_{i}$ has molecular area $a_{i}$, an adsorption term
including steric exclusion would be $\gamma
n_{s}(1-\sum_{i=1}^{N}a_{i}\G_{i})$.  The surface rate equations
remain linear in ${\bf \Gamma}$, but with a time-dependent transition
matrix ${\bf M}$. The effective boundary condition
Eq.\,\ref{EFFECTIVEBOUNDARY} is now nonlinear in $n_{s}(t)$ through
$F[n_{s}]$. However, for many biochemical applications (such as cell
signaling and virus adsorption and entry) the total surface
concentration is low such that $\sum_{i=1}^{N}a_{i}\G_{i} \ll 1$ and
the adsorption term can be linearized. In our one-dimensional
analysis, as long as $N$ is not too large and there is an appreciable
annihilation process, the surface concentrations all vanish in
time. 

The effects of multistage adsorption can also be explored on surfaces
of arbitrary shape, particularly for cylinders and spheres.  For
multistage processes on a sphere, the sublayer concentration
approaches a positive value $n_{s}(t\rightarrow \infty) = n_{0}(1-
{\gamma K_{0} \over D+\gamma K_{0}})$. We also expect positive
eigenvalues $\lambda_{j} >0$ of ${\bf M}$ to have a striking effect on
the transport. 


Finally, although we have only considered simple linearized surface
reaction schemes with negative eigenvalues, systems that support
oscillations, such as those involved in surface-mediated cell
signalling, could also be treated within our framework. Features of
the surface reactions and the bulk concentrations near the reacting
surface remain coupled through the kernel $K(t)$. Under certain
conditions, nonlinear surface reaction schemes may also be
linearized. One example is in the {\it stochastic} representation of
the surface reactions.  If we write the surface quantities in terms of
the probability distribution function $P(n_{1},n_{2},n_{3},\ldots,t)$
that there are $n_{1}$ molecules of of type 1, $n_{2}$ of type 2,
etc., the surface reactions can be written as a linear Master
equation. This allows our approach to be applied when $\Gamma_{1}$ in
the last term of Eq. \ref{NS} is interpreted as $\langle
\Gamma_{1}(t)\rangle$, the ensemble average
$\sum_{\{n_{i}\}}n_{1}P(\{n_{i}\},t)$. Using this interpretation, the
full problem can be solved using linear methods similar to those
presented, albeit for extremely large matrix dimension.

\vspace{2mm}

\noindent The authors thank Benhur Lee for stimulating
discussions.  TC acknowledges support from the NSF
(DMS-0349195), and the NIH (K25AI41935). Part of this work was
done during the Cells and Materials Workshop at IPAM UCLA.

\section{Appendix}

Here, we derive asymptotic expression for bulk and surface densities.
The trivial short time behavior of the subsurface density is
$n_{s}(t)/n_{0} \sim (1-\gamma t)$, independent of the surface
reactions since the first physical phenomenon to occur is particle
adsorption from the bulk to the interface, at rate $\gamma$.

For large distances $(z-1)/\sqrt{Dt} \gg 1$ and in the limit $\gamma
\tilde{K}(s=0)\equiv \gamma K_{0} \gg \sqrt{D/t}$, asymptotic
evaluation of the inverse Laplace inversion integral over
$\tilde{n}(z,s)$ yields

\begin{equation}
\begin{array}{l}
\displaystyle {n(z,t) \over n_{0}} \sim 1 - \left(1+{\sqrt{\pi D}
\over \gamma K_{0}\sqrt{t}}\right) \exp\left[-{(z-1)^{2} \over
4Dt}\right].
\label{LARGEZ}
\end{array}
\end{equation}

\noindent The condition $\gamma K_0 \gg \sqrt{D / t}$ can be
interpreted as a comparison between two typical velocities. The usual
diffusive velocity, $\sqrt{D/t}$, is compared to an effective reaction
velocity expressed by $\gamma$ modulated by surface effects through
the kernel $K_0$.  We may thus define an effective Damk\"ohler number
$D_a \equiv \gamma K_0 \sqrt{t} / \sqrt {D}$, so that
Eqn.\,\ref{LARGEZ} is valid only at large distances and for large
values of $D_a$.  The leading term on the right-hand-side above is
independent of the surface kinetics: the first information to have
traveled away from the interface is the initial depletion of the
$n_{s}$ layer into the surface and interfacial effects emerge as first
order corrections.

In the $t\rightarrow \infty$ limit, the dominant contribution to
$n(z,t)$ comes from small values of $u$ in Eq.\,\ref{NZT}.
Approximating $L(1,u)$ with its $u\rightarrow 0$ limit, we find the
asymptotic long time limit

\begin{equation}
\begin{array}{rl}
n_{s}(t) & \displaystyle \sim {n_{0} \over \gamma K_{0}}\sqrt{D \over \pi t}\\[13pt]
\: & \sim \displaystyle {n_0\over \sqrt{t}}\left[{\sqrt{D} \over
\gamma \sqrt{\pi} \left(1+\sum_{j=1}^{N}V^{-1}_{1j}V_{j1}\lambda_{j}^{-1}\right)}\right].
\label{LONGTIME}
\end{array}
\end{equation}

\noindent This expression is valid only if the surface dynamics
include a net sink of material. As long as there is some annihilation,
$K_{0}= 1+\sum_{j=1}^{N}V^{-1}_{1j}V_{j1}/\lambda_{j} < 0$ and
Eq.\,\ref{LONGTIME} holds.  A similar consideration of the small-$u$
dominated integration in Eq.\,\ref{GAMMA2} yields for the surface
concentrations

\begin{equation}
\displaystyle \Gamma_{k}(\vert\lambda^{*}\vert t\rightarrow \infty)
\sim {n_{0} \over K_{0}}\sqrt{D \over \pi t}\sum_{j=1}^{N}
V^{-1}_{kj}V_{j1} \vert\lambda_{j}\vert^{-1},
\label{GAMMAASYMP}
\end{equation}

\noindent where $\lambda_{*}<0$ is the largest eigenvalue of the
chemical transition matrix ${\bf M}$. Both Eq. \ref{LONGTIME} and
\ref{GAMMAASYMP} exhibit diffusion-limited $1/\sqrt{t}$ behavior.
These general results rely only on the linearity of $F[n_{s}]$ and are
valid for {\it any} surface reaction scheme through the eigenvalues
and eigenvectors of the transition matrix ${\bf M}$ and the resulting
function $\tilde{K}(s)$.


\end{document}